\documentclass[12pt]{article}
\usepackage{amssymb}

\usepackage{graphicx}

\ifx\pdfoutput\undefined
    \relax
\else
    \usepackage{epstopdf}
\fi

\makeatletter
\def\numberbysection{\@addtoreset{equation}{section}
 	\def\theequation{\thesection.\arabic{equation}}}
\makeatother

\numberbysection

\newcommand{\be}{\begin{eqnarray}}
\newcommand{\ee}{\end{eqnarray}}
\newcommand{\non}{\nonumber}

\newcommand{\ch}{\mathop{\rm cosh}\nolimits}
\newcommand{\sh}{\mathop{\rm sinh}\nolimits}
\newcommand{\tnh}{\mathop{\rm tanh}\nolimits}
\newcommand{\cth}{\mathop{\rm coth}\nolimits}
\newcommand{\csch}{\mathop{\rm cosech}\nolimits}
\newcommand{\sech}{\mathop{\rm sech}\nolimits}
\newcommand{\h}{\ensuremath{\mathsf{h}}}
\newcommand{\sgn}{\mathop{\rm sgn}\nolimits}

\newcommand{\hh}{\mathop{\mathcal H}\nolimits}

\begin{document}

\begin{titlepage}
\strut\hfill UMTG--252
\vspace{.5in}
\begin{center}

\LARGE Boundary energy of the general open XXZ chain\\
\LARGE at roots of unity\\[1.0in]
\large Rajan Murgan, Rafael I. Nepomechie and Chi Shi\\[0.8in]
\large Physics Department, P.O. Box 248046, University of Miami\\[0.2in]  
\large Coral Gables, FL 33124 USA\\

\end{center}

\vspace{.5in}

\begin{abstract}
 We have recently proposed a Bethe Ansatz solution of the open
 spin-$1/2$ XXZ quantum spin chain with general integrable boundary
 terms (containing six free boundary parameters) at roots of
 unity.  We use this solution, together with an appropriate string
 hypothesis, to compute the boundary energy of the chain in
 the thermodynamic limit.
\end{abstract}
\end{titlepage}

\setcounter{footnote}{0}

\section{Introduction}\label{sec:intro}

There has been considerable interest in the open spin-1/2 XXZ
quantum spin chain with general integrable boundary terms
\cite{dVGR,GZ}, whose Hamiltonian can be written as
\be
\hh &=& {1\over 2}\sum_{n=1}^{N-1}\left( 
\sigma_{n}^{x}\sigma_{n+1}^{x}+\sigma_{n}^{y}\sigma_{n+1}^{y} 
+\ch \eta\ \sigma_{n}^{z}\sigma_{n+1}^{z}\right) \non \\
&+& {1\over 2}\sh \eta \Big[ 
\cth \alpha_{-} \tnh \beta_{-}\sigma_{1}^{z}
+ \csch \alpha_{-} \sech \beta_{-}\big( 
\ch \theta_{-}\sigma_{1}^{x} 
+ i\sh \theta_{-}\sigma_{1}^{y} \big) \non \\
& & \quad -\cth \alpha_{+} \tnh \beta_{+} \sigma_{N}^{z}
+ \csch \alpha_{+} \sech \beta_{+}\big( 
\ch \theta_{+}\sigma_{N}^{x}
+ i\sh \theta_{+}\sigma_{N}^{y} \big)
\Big]  \,, \label{Hamiltonian} 
\ee
where $\eta$ is the bulk anisotropy parameter, and
$\alpha_{\pm}\,, \beta_{\pm}\,, \theta_{\pm}$ are free boundary 
parameters. \footnote{Under a global 
spin rotation about the $z$ axis, the bulk terms  
remain invariant, and the boundary parameters $\theta_{\pm}$ become 
shifted by the same constant, $\theta_{\pm}\mapsto \theta_{\pm} + const$.
Hence, the energy (and in fact, the transfer matrix eigenvalues) 
depend on $\theta_{\pm}$ only through the difference $\theta_{-} - 
\theta_{+}$.}

Except for the special case $\alpha_{\pm}$ or $\beta_{\pm} \rightarrow
\infty$ when the boundary terms become diagonal \cite{Ga, ABBBQ, Sk},
the boundary terms break the bulk $U(1)$ symmetry generated by
$S^{z}$; i.e., the model has no continuous symmetry.  For generic 
values of boundary parameters, this model does
not seem to have a simple pseudovacuum, which precludes constructing
a conventional algebraic Bethe Ansatz solution.  Being associated with
the spin-1/2 representation of $U_{q}(su(2))$, this model is but the
simplest of an infinite hierarchy of more complicated integrable
quantum spin chains involving higher-dimensional representations
and/or higher-rank algebras.  Hence, solving the former model is
presumably a prerequisite for solving any of the latter ones.  This
model also has numerous applications in statistical mechanics,
condensed matter and quantum field theory.

A Bethe Ansatz solution of this model was found in \cite{CLSW}-\cite{YNZ} 
for the case that the boundary parameters obey the constraint
\be
\alpha_- + \epsilon_1 \beta_- + \epsilon_2 \alpha_+ + 
\epsilon_3 \beta_+ = \epsilon_0 (\theta_- -\theta_+) 
+\eta k + \frac{1-\epsilon_2}{2}i\pi \quad {\rm mod}\, (2i\pi) \,,
\quad \epsilon_1 \epsilon_2 \epsilon_3=+1 \,, 
\label{constraint}
\ee
where $\epsilon_i=\pm 1$, and $k$ is an integer such that $|k|\leq
N-1$ and $N-1+k$ is even.  Completeness of this solution is not
straightforward, as two sets of Bethe Ansatz equations are generally
needed in order to obtain all $2^{N}$ levels \cite{NR}. Related
work includes \cite{Do}-\cite{Baj}.
 
There remained the problem of solving the model (\ref{Hamiltonian})
when the constraint (\ref{constraint}) is {\it not} satisfied.
Building on earlier work \cite{MN1, MN2}, we recently proposed in
\cite{MNS} a solution of the model for arbitrary values of the
boundary parameters, provided that the bulk anisotropy parameter has
values
\be
\eta = {i \pi\over p+1}\,, 
\label{etavalues}
\ee
where $p$ is a positive integer. Hence, $q \equiv e^{\eta}$ is a
root of unity, satisfying $q^{p+1}=-1$.

As is well known, for both the closed chain and the open chain with 
diagonal boundary terms, the eigenvalues of the 
Hamiltonian (and more generally, the transfer matrix) can be 
expressed in terms of zeros (``Bethe roots'') of a single function 
$Q(u)$. This is in sharp contrast with the solution \cite{MNS}, which 
involves multiple $Q$ functions, and therefore, multiple sets of 
Bethe roots. The number of such $Q$ functions depends on the value of 
$p$. (Generalized $T-Q$ equations involving two such $Q$ functions 
first arose in \cite{MN2} for special values of the boundary 
parameters.)

The solution \cite{MNS} has additional properties which distinguish it
from typical Bethe Ansatz solutions: the $Q$ functions also have
normalization constants which must be determined; and the
Bethe Ansatz equations have a nonconventional form.  Given the
unusual nature of this solution, one can justifiably wonder
whether it provides a practical means of computing properties of
the chain in the thermodynamic ($N \rightarrow \infty$) limit.  To
address this question, we set out to compute the so-called boundary 
or surface energy (i.e., the order 1 contribution to the ground-state energy),
which is perhaps the most accessible boundary-dependent quantity.
For the case of diagonal boundary terms, this quantity was first 
computed numerically in  \cite{ABBBQ}, and then analytically in 
\cite{HQB}.

We find that the boundary energy computation is indeed feasible.  The
key point is that, when the boundary parameters are in some suitable
domain, the ground-state Bethe roots appear to follow certain
remarkable patterns.  By assuming the strict validity of these
patterns (``string hypothesis''), the Bethe equations reduce to a
conventional form.  Hence, standard techniques can then be used to
complete the computation.  We find that our final result
(\ref{XXZboundenergyeach}) for the boundary energy coincides with the
result obtained in \cite{finitesize} for the case that the boundary
parameters obey the constraint (\ref{constraint}), and in \cite{MNS2}
for special values \cite{MN1, MN2} of the boundary parameters at 
roots of unity.

The outline of this paper is as follows.  In Section \ref{sec:bethe},
we briefly review the Bethe Ansatz solution \cite{MNS} of the model
(\ref{Hamiltonian}) at roots of unity (\ref{etavalues}).  In Section
\ref{sec:even}, we treat the case of even $p$, followed by the case of
odd $p$ in Section \ref{sec:odd}.  This is followed by discussion of our
results and a brief outline of our future work in Section \ref{sec:conc}.
    
\section{Bethe Ansatz}\label{sec:bethe}

In this section, we briefly recall the Bethe Ansatz solution
\cite{MNS}.  In order to ensure hermiticity of the Hamiltonian (1.1),
we take the boundary parameters $\beta_{\pm}$ real; $\alpha_{\pm}$
imaginary; $\theta_{\pm}$ imaginary.  We begin by introducing the
Ansatz for the various $Q(u)$ functions that appear in our solution, which
we denote as $a_{j}(u)$ and $b_{j}(u)$:
\be
a_{j}(u) &=& A_{j} \prod_{k=1}^{2M_{a}} \sinh(u-u_{k}^{(a_{j})}) \,, \qquad 
b_{j}(u) = B_{j}\prod_{k=1}^{2M_{b}} \sinh(u-u_{k}^{(b_{j})})\,, \non \\
& & \qquad j = 1\,, \ldots \,, \lfloor{p+1\over 2}\rfloor \,, \label{Ansatzpgen2} 
\ee
where $\{ u_{k}^{(a_{j})} \,, u_{k}^{(b_{j})} \}$ are the zeros of $a_{j}(u)$ 
and $b_{j}(u)$ respectively, and $\lfloor{\ }\rfloor$ denotes integer part.
If $p$ is even, then there is one additional set of functions 
corresponding to $j=\frac{p}{2}+1$,
\be 
a_{{p\over 2}+1}(u) &=& A_{{p\over 2}+1} 
\prod_{k=1}^{M_{a}} \sinh(u-u_{k}^{(a_{{p\over 2}+1})})
\sinh(u+u_{k}^{(a_{{p\over 2}+1})}) \,, \non \\
b_{{p\over 2}+1}(u) &=& B_{{p\over 2}+1}
\prod_{k=1}^{M_{b}} \sinh(u-u_{k}^{(b_{{p\over 2}+1})}) 
\sinh(u+u_{k}^{(b_{{p\over 2}+1})}) \,.
\label{Ansatzcross}
\ee 
The normalization constants $\{ A_{j}\,, B_{j} \}$ are yet to be
determined \footnote{One of these normalization constants can be set
to unity.}.  We assume that $N$ is even, in which case the integers
$M_{a}\,, M_{b}$ are given by
\be
M_{a} = {N\over 2} + 2p \,, \qquad 
M_{b} = {N\over 2} + p - 1 \,,
\label{Mvaluespgen}
\ee
It is clear from (\ref{Ansatzpgen2}), (\ref{Ansatzcross}) that
$a_{j}(u)$ and $b_{j}(u)$ have the following periodicity and crossing
properties,
\be
a_{j}(u+ i\pi) &=& a_{j}(u) \,, \qquad \quad b_{j}(u+ i\pi) = b_{j}(u) \,, 
\qquad j = 1\,, \ldots \,, \lfloor{p\over 2}\rfloor+1 \,, \label{periodicity} \\
a_{{p\over 2}+1}(-u) &=& a_{{p\over 2}+1}(u) \,, \qquad
b_{{p\over 2}+1}(-u) = b_{{p\over 2}+1}(u) \,. \label{crossing}
\ee 

The zeros of the functions $\{ a_{j}(u)\}$ and $\{ b_{j}(u)\}$ satisfy the
following Bethe Ansatz equations
\be
{h_{0}(-u_{l}^{(a_{1})}-\eta)\over h_{0}(u_{l}^{(a_{1})})}
&=&-{f_{1}(u_{l}^{(a_{1})})\ a_{1}(-u_{l}^{(a_{1})}) +
g_{1}(u_{l}^{(a_{1})})\ Y(u_{l}^{(a_{1})})^{2}\ 
b_{1}(-u_{l}^{(a_{1})})\over
2a_{2}(u_{l}^{(a_{1})})\ h_{1}(-u_{l}^{(a_{1})}-\eta)\
\prod_{k=1}^{p}h_{1}(u_{l}^{(a_{1})}+k\eta)} \,, \label{BAEpgena1} \\
{h(-u_{l}^{(a_{j})}-j\eta)\over 
h(u_{l}^{(a_{j})}+(j-1)\eta)}&=&-{a_{j-1}(u_{l}^{(a_{j})})\over
a_{j+1}(u_{l}^{(a_{j})})}  \,, \qquad 
j = 2 \,, \ldots \,, \lfloor{p\over 2}\rfloor+1 \,,
\label{BAEpgena}
\ee 
and 
\be
{h_{0}(-u_{l}^{(b_{1})}-\eta)\over h_{0}(u_{l}^{(b_{1})})}
&=&-{f_{1}(u_{l}^{(b_{1})})\ b_{1}(-u_{l}^{(b_{1})}) +
g_{1}(u_{l}^{(b_{1})})\  
a_{1}(-u_{l}^{(b_{1})})\over
2b_{2}(u_{l}^{(b_{1})})\ h_{1}(-u_{l}^{(b_{1})}-\eta)\
\prod_{k=1}^{p}h_{1}(u_{l}^{(b_{1})}+k\eta)} \,, \label{BAEpgenb1} \\
{h(-u_{l}^{(b_{j})}-j\eta)\over 
h(u_{l}^{(b_{j})}+(j-1)\eta)}&=&-{b_{j-1}(u_{l}^{(b_{j})})\over
b_{j+1}(u_{l}^{(b_{j})})}  \,, \qquad 
j = 2 \,, \ldots \,, \lfloor{p\over 2}\rfloor+1 \,,
\label{BAEpgenb}
\ee 
where $a_{{p\over2}+2}(u)=a_{{p\over2}}(-u)$ and
$a_{{p+3\over2}}(u)=a_{{p+1\over2}}(-u)$ for even and odd values of
$p$, respectively, and similarly for the $b$'s.
Moreover,
\be
h(u) = h_{0}(u)\ h_{1}(u) \,,
\label{hfunction}
\ee
where $h_{0}(u)$ and $h_{1}(u)$ are as follows
\be
h_{0}(u) &=& \sinh^{2N}(u+\eta){\sinh(2u+2\eta)\over \sinh(2u+\eta)} 
\,, \non  \\
h_{1}(u) &=& - 4 \sinh(u+\alpha_{-}) \cosh(u+ \beta_{-}) 
\sinh(u+ \alpha_{+}) \cosh(u+ \beta_{+}) \,. 
\ee
We also define the quantities
\be
g_{1}(u) = 2 \sinh (2(p+1)u) 
\label{g}
\ee
and 
\be 
Y(u)^2 = \sum_{k=0}^{2} \mu_{k} \cosh^{k}(2(p+1)u) \,.
\label{Y}
\ee 
Explicit expressions for the coefficients $\mu_{k}$ in (\ref{Y}),
which depend on the boundary parameters, as well as for the function 
$f_{1}(u)$, are listed in the Appendix for both even and odd values of $p$.

Moreover, there are additional Bethe-Ansatz-like equations 
\be
a_{1}({\eta\over 2}) &=& a_{2}(-{\eta\over 2}) \,,
\label{newbaea1} \\
a_{j-1}(({1\over 2}-j)\eta) &=&  a_{j+1}(({1\over 2}-j)\eta) \,, 
\qquad j = 2 \,, \ldots \,, \lfloor{p\over 2}\rfloor +1 \,,
\label{newbaeaj}
\ee 
which relate the normalization constants $\{ A_{j} \}$; and also 
\be
b_{1}({\eta\over 2}) &=& b_{2}(-{\eta\over 2}) \,,
\label{newbaeb1}\\
b_{j-1}(({1\over 2}-j)\eta) &=&  b_{j+1}(({1\over 2}-j)\eta) \,, 
\qquad j = 2 \,, \ldots \,, \lfloor{p\over 2}\rfloor +1 \,, 
\label{newbaebj}
\ee 
which relate the normalization constants $\{ B_{j} \}$. There are also
equations that relate the normalization constants $A_{1}$ and $B_{1}$, such as 
\be
f_{1}(-\alpha_{-}-\eta)\ b_{1}(\alpha_{-}+\eta) =
-g_{1}(-\alpha_{-}-\eta)\  a_{1}(\alpha_{-}+\eta)
\,.
\label{BAEnorm}
\ee

The energy eigenvalues of the Hamiltonian (\ref{Hamiltonian}) are 
given by
\be
E &=& {1\over 2}\sinh \eta \sum_{l=1}^{2M_{b}} \left[
\coth (u_{l}^{(b_{j})}+(j-1)\eta) - \coth 
(u_{l}^{(b_{j-1})}+(j-1)\eta) \right] +E_{0} \,, \non \\
& & \qquad j= 2\,, \ldots \,, \lfloor{p+1\over 2}\rfloor
\,, \label{energypgen}
\ee
where $E_{0}$ is defined as
\be
E_{0} = {1\over 2}\sinh \eta \left( \coth \alpha_{-} + \tanh \beta_{-} 
+ \coth \alpha_{+} + \tanh \beta_{+} \right) 
+ {1\over 2}(N-1)\cosh \eta \,. \label{E0}
\ee
For even $p$, there is one more expression for the energy 
corresponding to $j=\frac{p}{2}+1$,
\be
E &=& {1\over 2}\sinh \eta  \Bigg\{ \sum_{l=1}^{M_{b}} \left[
\coth (u_{l}^{(b_{{p\over 2}+1})}+{p\eta\over 2}) - 
\coth (u_{l}^{(b_{{p\over 2}+1})}-{p\eta\over 2}) \right] \non \\
&-& \sum_{l=1}^{2M_{b}} \coth (u_{l}^{(b_{p\over 2})}+{p\eta\over 2}) 
\Bigg\}  + E_{0} \,.
\label{energyspecial}
\ee 
There are also similar expressions for the energy in terms of $a$ 
roots $\{ u_{l}^{(a_{j})}\}$ \cite{MNS}.

\section{Even $p$}\label{sec:even}

\begin{figure}[htb]
	\centering
	\includegraphics[width=0.80\textwidth]{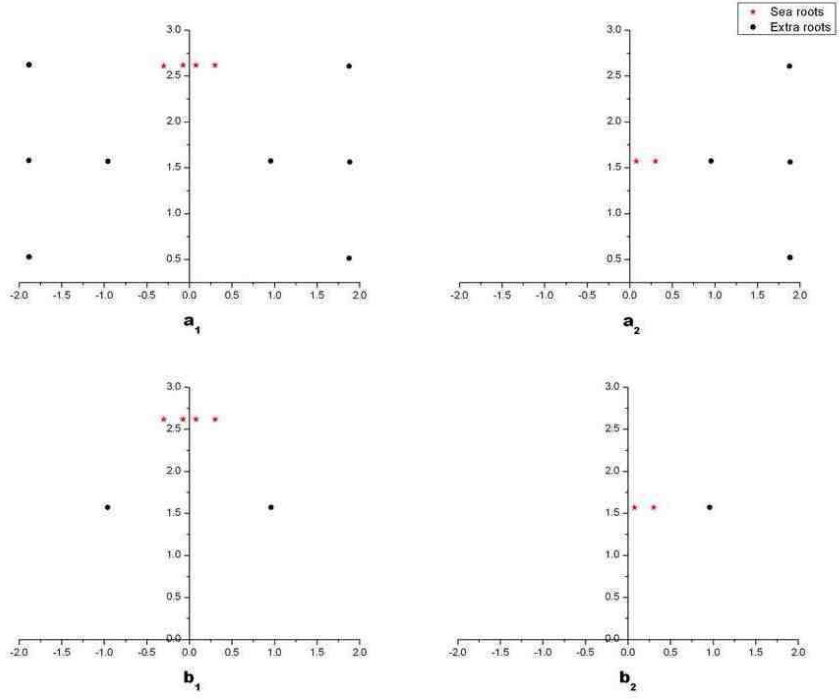}
	\caption[xxx]{\parbox[t]{0.8\textwidth}{
Ground-state Bethe roots for $p=2$, $N=4$, $\alpha_{-}=0.604i$, 
$\alpha_{+}=0.535i$, $\beta_{-}=-1.882$, $\beta_{+}=1.878$, 
$\theta_{-}=0.6i$, $\theta_{+}=0.7i$.}
	}
	\label{fig:p2}
\end{figure}

In this section, we consider the case where the bulk anisotropy parameter
assumes the values (\ref{etavalues}) with $p$ even, i.e., 
$\eta = {i\pi\over 3}\,, {i\pi\over 5}\,, \ldots$.
We have studied the Bethe roots corresponding to 
the ground state numerically for small values of $p$ and $N$ 
along the lines of \cite{NR}.
We have found that, when the boundary parameters are in some
suitable domain (which we discuss further below Eq. (\ref{XXZboundenergyeach})), 
the ground state Bethe roots $\{u_{k}^{(a_{j})}, u_{k}^{(b_{j})}\}$ 
have a remarkable pattern. An example with $p=2\,, N=4$ is shown in Figure \ref{fig:p2}.
Specifically, these roots can be categorized into ``sea'' roots, 
$\{v_{k}^{\pm (a_{j})}, v_{k}^{\pm (b_{j})}\}$ (the number of which depends on $N$) 
and the remaining ``extra'' roots, $\{w_{k}^{\pm (a_{j},l)}, w_{k}^{\pm (b_{j})}\}$ 
(the number of which depends on $p$)
according to the following pattern which we adopt as our ``string hypothesis''.

\subsection{Sea roots $\{v_{k}^{\pm (a_{j})}\,, v_{k}^{\pm 
(b_{j})}\}$} \label{subsec:evensearoots}

Sea roots of all $\{a_{j}(u) \,, b_{j}(u) \}$ functions for any even $p$ 
are summarized below,
\be
v_{k}^{\pm (a_{j})} &=& v_{k}^{\pm (b_{j})} = 
\pm \tilde v_{k} + \left({2 p + 3 - 2 j\over 2}\right)\eta\,,
\quad k = 1\,,\ldots\,, {N\over 2} \,, \non \\
 & &\qquad j = 1\,,\ldots \,, {p\over 2} + 1 \,, \label{seapeven}
\ee   
where $\tilde v_{k}$ are real and positive. In Figure \ref{fig:p2}, the sea 
roots are indicated with red stars.

Note that the real parts ($\pm \tilde v_{k}$) are independent of $j$. 
This, as we shall see, greatly simplifies the analysis. Furthermore, 
for each sea root with real part $+\tilde v_{k}$, there is an 
additional ``mirror'' sea root with real part $-\tilde v_{k}$, for a 
total of $N$ sea roots, provided $j \ne \frac{p}{2}+1$. For 
$j = \frac{p}{2}+1$, there are only $\frac{N}{2}$ sea roots $+\tilde v_{k}+ 
\frac{i \pi}{2}$ (i.e., just the root with positive real part) due to the 
crossing symmetry (\ref{crossing}) of the functions $a_{{p\over2}+1}(u)$ and 
$b_{{p\over2}+1}(u)$. \footnote{Hence, strictly speaking, we should 
write the $j = \frac{p}{2}+1$ equation in (\ref{seapeven}) 
separately, keeping only the $+$ roots. However, in order to avoid 
doubling the number of equations, we commit this abuse of notation 
here and throughout this section.} 

\subsection{Extra roots $\{ w_{k}^{\pm (a_{j},l)}\,, w_{k}^{\pm 
(b_{j})}\}$} \label{subsec:evenextraroots}

We next describe the remaining extra Bethe roots for even $p$, the number
of which depends on the value of $p$. In Figure \ref{fig:p2}, the 
extra roots are indicated with black circles.
Since the functions $a_{j}(u)$ and $b_{j}(u)$ have a different number
of such extra roots, we present them separately.  The extra roots of
the $b_{j}(u)$ functions have the form
\be
w_{k}^{\pm (b_{j})} &=& \pm \tilde w_{k} 
+ \left({2 p + 1 - 2 k\over 2}\right)\eta\,,
\quad k = 1\,,\ldots\,, p - 1 \,, \non\\
& & \qquad j = 1\,,\ldots \,, {p\over 2} + 1 \,. \label{extrabpeven}
\ee  
The real parts of the roots, $\tilde w_{k}$, are not all independent. 
Instead, they are related to each other pairwise as follows,
\be
\tilde w_{k}=\tilde w_{p - k}\,, \qquad k = 1\,,\ldots \,, {p\over 2} 
- 1 \,. \label{wkrltn}
\ee
Only $\tilde w_{{p\over2}}$ remains unpaired. 
This property proves to be crucial for the boundary energy calculation. 

There are two types of extra roots of the $a_{j}(u)$ functions:
\be
w_{k}^{\pm (a_{j},1)} &=& w_{k}^{\pm (b_{j})} =
\pm \tilde w_{k} + \left({2 p + 1 - 2 k\over 2}\right)\eta\,,
\quad k = 1\,,\ldots \,, p - 1 \,, \non\\
w_{k}^{\pm (a_{j},2)} &=& \pm \tilde w_{0} 
+ \left({2 p + 3 - 2 k\over 2}\right)\eta\,,
\quad k = 1\,,\ldots \,, p + 1 \,, \non \\
& & \qquad j = 1\,,\ldots \,, {p\over 2} + 1 \,.
\ee  
Note that the extra roots of the first type $\{ w_{k}^{\pm (a_{j},1)} \}$ 
coincide with the $b$ roots $\{ w_{k}^{\pm (b_{j})} \}$;
and that the extra roots of the second type
$\{ w_{k}^{\pm (a_{j},2)} \}$ form a ``$(p + 1)$-string'', with real part 
$\tilde w_{0}$. 

As previously remarked, for $j = {p\over2} + 1$, only the roots with the $+$ sign appear.

\subsection{Boundary energy}

We now proceed to compute the boundary energy. Using the expression 
(\ref{energypgen}) for the energy and our string hypothesis, we 
obtain (for $p>2$)
\be
E &=& {1\over 2}\sinh \eta  \Bigg\{ \sum_{k=1}^{\frac{N}{2}} \Big[
\coth (v_{k}^{+(b_{j})}+(j-1)\eta) + \coth 
(v_{k}^{-(b_{j})}+(j-1)\eta) \non \\
&-&\coth (v_{k}^{+(b_{j-1})}+(j-1)\eta) - 
\coth (v_{k}^{-(b_{j-1})}+(j-1)\eta) \Big]\non \\
&+& \sum_{k=1}^{p-1} \Big[
\coth (w_{k}^{+(b_{j})}+(j-1)\eta) + \coth 
(w_{k}^{-(b_{j})}+(j-1)\eta) \non \\
&-&\coth (w_{k}^{+(b_{j-1})}+(j-1)\eta) - 
\coth (w_{k}^{-(b_{j-1})}+(j-1)\eta)\Big] 
\Bigg\} + E_{0}\,, \non \\
& & \qquad j= 2\,, \ldots \,, {p\over 2} \,.
\ee 
Recalling (\ref{seapeven}) and (\ref{extrabpeven}), 
this expression for the energy reduces to
\be
E &=& \sinh^{2}\eta \sum_{k=1}^{{N\over2}}{1\over 
\sinh (\tilde v_{k} - {\eta\over2})
\sinh (\tilde v_{k} + {\eta\over2})} + E_{0} \,,\quad \tilde v_{k} > 
0 \,,
\label{energyI1}
\ee 
independently of the value of $j$.
Since the extra roots $w_{k}^{(b_{j})}$ are independent of $j$, their
contribution to the energy evidently cancels, leaving only the sea-root 
terms in (\ref{energyI1}).  The same result can also be obtained (for 
$p \ge 2$) from the energy expression (\ref{energyspecial}).

We turn now to the Bethe Ansatz equations, on which we must also impose our
string hypothesis.  Choosing $j={p\over 2}+1$ in (\ref{BAEpgenb}) with
$u_{l}^{(b_{j})}$ equal to the sea root $v_{l}^{+(b_{{p\over
2}+1})}=\tilde v_{l}+ \frac{i\pi}{2}$, we obtain
\be
{h(-\tilde v_{l}-\frac{\eta}{2})\over 
h(\tilde v_{l}-\frac{\eta}{2})}&=&
-{b_{\frac{p}{2}}(\tilde v_{l}+\frac{i\pi}{2} )\over
b_{\frac{p}{2}}(-\tilde v_{l}-\frac{i\pi}{2})} \,,
\ee
where we have made use of the fact $b_{{p\over2}+2}(u)=b_{{p\over2}}(-u)$.
More explicitly, this equation reads
\be
\lefteqn{\left(
{\sinh(\tilde v_{l} + {\eta\over 2}) \over 
\sinh (\tilde v_{l} - {\eta\over 2})}\right)^{2N}
{\sinh(2\tilde v_{l} + \eta) \over 
\sinh (2\tilde v_{l} - \eta)}
{\sinh(\tilde v_{l} - {\eta\over 2} + \alpha_{-}) \over 
\sinh (\tilde v_{l} + {\eta\over 2} - \alpha_{-})}
{\cosh(\tilde v_{l} - {\eta\over 2} + \beta_{-}) \over 
\cosh (\tilde v_{l} + {\eta\over 2} - \beta_{-})}} \non \\
& & \times 
{\sinh(\tilde v_{l} - {\eta\over 2} + \alpha_{+}) \over 
\sinh (\tilde v_{l} + {\eta\over 2} - \alpha_{+})}
{\cosh(\tilde v_{l} - {\eta\over 2} + \beta_{+}) \over 
\cosh (\tilde v_{l} + {\eta\over 2} - \beta_{+})}
= - \prod_{k=1}^{N\over 2} 
{\sinh(\tilde v_{l} - \tilde v_{k} + \eta) \over 
\sinh (\tilde v_{l} - \tilde v_{k} - \eta)}
{\sinh(\tilde v_{l} + \tilde v_{k} + \eta) \over 
\sinh (\tilde v_{l} + \tilde v_{k} - \eta)} \,, \non \\
& & \qquad l = 1 \,, \cdots \,, {N\over 2} \,,\quad \tilde v_{k} > 0 \,.
\label{BAEpeven}
\ee
In obtaining this result, we have made use of the fact that the
normalization constant $B_{\frac{p}{2}}$ of the function
$b_{\frac{p}{2}}(u)$ cancels, and also that the contribution from the
extra roots on the RHS cancel as a consequence of the relation
(\ref{wkrltn}) among their real parts.

Remarkably, as a consequence of our string hypothesis, our
non-conventional Bethe Ansatz equations have reduced to a
conventional system (\ref{BAEpeven}), which can be analyzed by
standard methods. However, before proceeding further with this computation,
it is worth noting that the same equations can also be obtained starting 
from any $j>1$. To see this, we first 
observe that the $\{A_{j} \}$ normalization constants are all equal, and 
similarly for the $\{B_{j} \}$ normalization constants, 
\be
A_{1}=A_{2}=\ldots=A_{{p\over 2}+1}\,, \qquad
B_{1}=B_{2}=\ldots=B_{{p\over 2}+1}\,.
\ee 
This result follows from the Bethe-Ansatz-like equations 
(\ref{newbaea1})-(\ref{newbaebj}) and the string hypothesis. For 
example, using (\ref{seapeven}) and (\ref{extrabpeven}) in 
(\ref{newbaeb1}), and remembering the relation (\ref{wkrltn}) among the real parts 
of the extra roots, we obtain $B_{1}=B_{2}$. Hence, choosing
$u_{l}^{(b_{j})}$ in (\ref{BAEpgenb})
to be a sea root $v_{l}^{+(b_{j})}$ for any $j \in \{2\,, \ldots\,,  
\frac{p}{2}+1 \}$, we again arrive at (\ref{BAEpeven}). Moreover, in view 
of the identity
\be
{a_{j-1}(v_{l}^{+(a_{j})})\over a_{j+1}(v_{l}^{+(a_{j})})} =
{b_{j-1}(v_{l}^{+(b_{j})})\over b_{j+1}(v_{l}^{+(b_{j})})}\,, \qquad 
j = 2 \,, \ldots \,, {p\over 2}+1 \,,
\ee
where $v_{l}^{+(a_{j})} = v_{l}^{+(b_{j})}$ is a sea root,
the same result (\ref{BAEpeven}) can also be obtained from (\ref{BAEpgena}).
\footnote{Only the first set of Bethe equations (\ref{BAEpgena1}), 
(\ref{BAEpgenb1}) do not seem to reduce to (\ref{BAEpeven}).}

In the thermodynamic ($N\rightarrow\infty$) limit, the number of sea
roots becomes infinite.  The distribution of the real parts of these
roots $\{\tilde v_{k}\}$ can be represented by a density function,
which is computed from the counting function.  
To this end, following \cite{finitesize, MNS2} and references therein, we define
some basic quantities
\be
e_{n}(\lambda) =
{\sinh \mu \left( \lambda + {i n\over 2} \right) 
\over \sinh \mu \left( \lambda - {i n\over 2} \right) } \,, \qquad
g_{n}(\lambda) = e_{n}(\lambda \pm {i \pi \over 2 \mu})
= {\cosh \mu \left( \lambda + {i n\over 2} \right) 
\over \cosh \mu \left( \lambda - {i n\over 2} \right) } \,,
\ee
which allow us to rewrite the Bethe Ansatz equations (\ref{BAEpeven})
in a more compact form, 
\be 
e_{1}(\lambda_{l})^{2N+1}\  g_{1}(\lambda_{l})
{e_{2a_{-}-1}(\lambda_{l})\ e_{2a_{+}-1}(\lambda_{l})\over 
g_{1+2ib_{-}}(\lambda_{l})\ g_{1+2ib_{+}}(\lambda_{l})} &=& 
-\prod_{k=1}^{N\over 2} 
e_{2}(\lambda_{l}-\lambda_{k})\ 
e_{2}(\lambda_{l}+\lambda_{k})  \,, \non \\ 
& & l = 1 \,, \cdots \,, {N\over 2} \,,
\label{BAEcompact}
\ee
where we have set $\tilde v_{l} = \mu \lambda_{l}$,
$\eta = i \mu \,, 
\alpha_{\pm} = i \mu a_{\pm} \,, \beta_{\pm} = \mu b_{\pm}$. Note 
that the parameters $\mu$, $a_{\pm}$, $b_{\pm}$ are all real.

Taking the logarithm of (\ref{BAEcompact}), we obtain the desired
ground state counting function
\be
\h(\lambda)&=&{1\over 2\pi}\Big\{ (2N+1) q_{1}(\lambda) + r_{1}(\lambda)
+ q_{2a_{-}-1}(\lambda) -  r_{1+2ib_{-}}(\lambda)
+ q_{2a_{+}-1}(\lambda) -  r_{1+2ib_{+}}(\lambda) \non \\
&-& \sum_{k=1}^{N\over 2} \left[ q_{2}(\lambda - \lambda_{k}) +
q_{2}(\lambda + \lambda_{k})  \right] \Big\} \,,
\label{counting}
\ee
where $q_{n}(\lambda)$ and $r_{n}(\lambda)$ are odd functions defined
by
\be
q_{n}(\lambda) &=& \pi + i \ln e_{n}(\lambda) 
= 2 \tan^{-1}\left( \cot(n \mu/ 2) \tanh( \mu \lambda) \right)
\,, \non \\
r_{n}(\lambda) &=&  i \ln g_{n}(\lambda) \,.
\label{logfuncts}
\ee
Defining $\lambda_{-k} \equiv -\lambda_{k}$, we have
\be
-\sum_{k=1}^{N\over 2} \left[ q_{2}(\lambda - \lambda_{k}) +
q_{2}(\lambda + \lambda_{k})  \right] =
-\sum_{k=-{N\over 2}}^{N\over 2} q_{2}(\lambda - \lambda_{k}) 
+ q_{2}(\lambda) \,.
\ee
The root density $\rho(\lambda)$ for the ground state is 
therefore given by
\be
\rho(\lambda) &=& {1\over N} {d \h\over d\lambda} 
 = 2 a_{1}(\lambda)
 - \int_{-\infty}^{\infty} d\lambda'\ a_{2}(\lambda - \lambda')\
 \rho(\lambda') +{1\over N}\Big[ a_{1}(\lambda) + b_{1}(\lambda) \non \\ 
 &+& a_{2}(\lambda) 
+ a_{2a_{-}-1}(\lambda) -  b_{1+2ib_{-}}(\lambda)
+ a_{2a_{+}-1}(\lambda) -  b_{1+2ib_{+}}(\lambda) \Big] \,, \label{lie}
\ee
where we have ignored corrections of higher order in $1/N$ when
passing from a sum to an integral, and we have introduced the
notations \footnote{These new functions $a_{n}(\lambda)$ and 
$b_{n}(\lambda)$ should not be confused with the $Q$ functions 
$a_{j}(u)$ and $b_{j}(u)$ appearing earlier. We apologize for this 
unfortunate coincidence of notations.}
\be
a_n(\lambda) &=& {1\over 2\pi} {d \over d\lambda} q_n (\lambda)
= {\mu \over \pi} 
{\sin (n \mu)\over \cosh(2 \mu \lambda) - \cos (n \mu)} \,, \non \\
b_n(\lambda) &=& {1\over 2\pi} {d \over d\lambda} r_n (\lambda)
= -{\mu \over \pi} 
{\sin (n \mu)\over \cosh(2 \mu \lambda) + \cos (n \mu)} \,. 
\ee 
The solution of the linear integral equation (\ref{lie}) for $\rho(\lambda)$
is obtained by Fourier transforms and is given by \footnote{Our 
conventions are
\be
\hat f(\omega) \equiv \int_{-\infty}^\infty e^{i \omega \lambda}\ 
f(\lambda)\ d\lambda \,, \qquad\qquad
f(\lambda) = {1\over 2\pi} \int_{-\infty}^\infty e^{-i \omega \lambda}\ 
\hat f(\omega)\ d\omega \,. \non 
\ee} 
\be
\rho(\lambda) = 2 s(\lambda) + {1\over N} R(\lambda) \,,
\label{rho}
\ee
where
\be
s(\lambda) = {1\over 2\pi} \int_{-\infty}^{\infty} d\omega\ 
e^{-i \omega \lambda} {1\over 2 \cosh(\omega/2)} 
= {1\over 2 \cosh (\pi \lambda)} \,,
\ee
and 
\be
\hat R(\omega) &=& {1\over{\left(1+\hat a_{2}(\omega) \right)}} 
\Big\{ \hat a_{1}(\omega) + \hat b_{1}(\omega) +  \hat a_{2}(\omega)
  - \hat b_{1+2ib_{-}}(\omega) -  \hat b_{1+2ib_{+}}(\omega) \non \\
 &+&  \hat a_{2a_{-}-1}(\omega) +  \hat a_{2a_{+}-1}(\omega) 
   \Big\} \,,
\ee
with
\be
\hat a_{n}(\omega) &=& \sgn(n) {\sinh \left( (\nu  - |n|) 
\omega / 2 \right) \over
\sinh \left( \nu \omega / 2 \right)} \,,
\qquad 0 \le |n| < 2 \nu  \,, \label{fourier1} \\
\hat b_{n}(\omega) &=&
-{\sinh \left( n \omega / 2 \right) \over
\sinh \left( \nu \omega / 2 \right)} \,,
\qquad \qquad\qquad\quad  0 < \Re e\ n < \nu  \,,
\label{fourier2}
\ee
where $\nu\equiv \frac{\pi}{\mu}=p+1$.  

Expressing the energy expression (\ref{energyI1}) in terms of the
newly defined quantities and letting $N$ become large, we obtain
\be
E &=& - {2\pi \sin \mu\over \mu} \sum_{k=1}^{\frac{N}{2}} a_{1}(\lambda_{k}) + E_{0} 
= - {\pi \sin \mu\over \mu} \Big\{ 
\sum_{k=-\frac{N}{2}}^{\frac{N}{2}} a_{1}(\lambda_{k}) - a_{1}(0)\Big\} + E_{0} \non\\
&=& - {\pi \sin \mu\over \mu} \Big\{ 
N\int_{-\infty}^{\infty}d\lambda\ a_{1}(\lambda)\ \rho(\lambda) - a_{1}(0) 
\Big\}
+ {1\over 2}(N-1) \cos \mu \non\\
&+& {1\over 2} \sin \mu \left( \cot \mu a_{-} + i\tanh \mu b_{-} +
\cot \mu a_{+} + i\tanh \mu b_{+} \right)\,,
\ee 
where again we ignore corrections that are higher order in $1/N$.
Substituting the result (\ref{rho}) for the root density, we obtain
\be
E = E_{bulk} + E_{boundary} \,,
\ee
where the bulk (order $N$) energy is given by
\be
E_{bulk} &=& - {2N \pi \sin \mu\over \mu} 
\int_{-\infty}^{\infty}d\lambda\ a_{1}(\lambda)\ s(\lambda) 
+ {1\over 2}N \cos \mu \non \\
&=&  - N \sin^{2} \mu \int_{-\infty}^{\infty}
d\lambda\ {1\over \left[\cosh(2 \mu \lambda) - \cos \mu \right] 
\cosh (\pi \lambda)} +  {1\over 2}N \cos \mu \,,
\label{bulkenergy}
\ee
which agrees with the well-known result \cite{YY}.  The
boundary (order $1$) energy is given by
\be
E_{boundary} = - {\pi \sin \mu\over \mu} I
 -{1\over 2}\cos \mu + {1\over 2} \sin \mu \left( 
\cot \mu a_{-} + i\tanh \mu b_{-} +
\cot \mu a_{+} + i\tanh \mu b_{+} \right)\,,
\label{boundenergyI}
\ee
where $I$ is the integral 
\be
I &=& \int_{-\infty}^{\infty}d\lambda\ a_{1}(\lambda) \left[
R(\lambda) - \delta(\lambda) \right] 
= {1\over 2\pi}  \int_{-\infty}^{\infty} d\omega\ \hat a_{1}(\omega)
\left[ \hat R(\omega) - 1 \right] \non \\
&=& 
 {1\over 2\pi}  \int_{-\infty}^{\infty} d\omega\ \hat s(\omega)
\Big\{ \hat a_{1}(\omega) + \hat b_{1}(\omega)  - 
1 \non \\
& & -  \hat b_{1+2ib_{-}}(\omega) - \hat b_{1+2ib_{+}}(\omega) 
 + \hat a_{2a_{-}-1}(\omega) +  \hat a_{2a_{+}-1}(\omega)  
 \Big\} \,. \label{integralI}
\ee
We further write the boundary energy as the sum of contributions
from the left and right boundaries, 
$E_{boundary}= E_{boundary}^{-} + E_{boundary}^{+}$.
The energy contribution from each boundary is given by
\be
E_{boundary}^{\pm} &=& - {\sin \mu\over 2\mu} 
\int_{-\infty}^{\infty} d\omega\ 
{1\over 2\cosh (\omega/ 2)}
\Big\{ 
{\sinh((\nu-2)\omega/4) \over 2\sinh(\nu \omega/4)} 
-{1\over 2} \non \\
&+& \sgn(2a_{\pm}-1){\sinh((\nu-|2a_{\pm}-1|)\omega/2) \over 
\sinh(\nu\omega/2)} + 
{\sinh((2ib_{\pm}+1)\omega/2)  \over 
\sinh(\nu \omega/2)} \Big\} \non \\
&+& {1\over 2} \sin \mu \left( \cot \mu a_{\pm} + i\tanh \mu b_{\pm}\right) 
 -{1\over 4}\cos \mu \,.
\label{XXZboundenergyeach}
\ee
This result can be shown to coincide with previous results in
\cite{finitesize, MNS2}.

\begin{figure}[htb]
	\centering
	\includegraphics[width=0.40\textwidth]{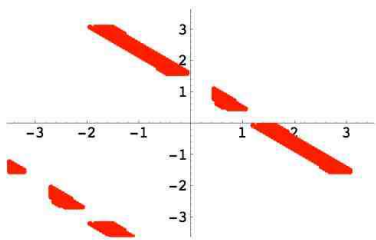}
	\caption[xxx]{\parbox[t]{0.8\textwidth}{
Shaded area denotes region of the $(\Im m\, \alpha_{+} \,, \Im m\,  
\alpha_{-} )$ plane for which the ground-state Bethe roots obey
the string hypothesis for $p=2$, $N=2$, $\beta_{-}=-1.882$, $\beta_{+}=1.878$, 
$\theta_{-}=0.6i$, $\theta_{+}=0.7i$.}
	}
	\label{fig:p2alpha}
\end{figure}

\begin{figure}[htb]
	\centering
	\includegraphics[width=0.40\textwidth]{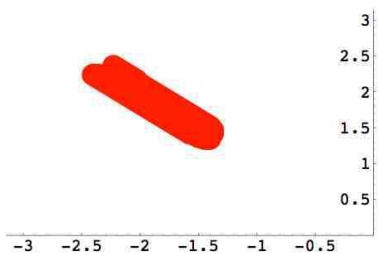}
	\caption[xxx]{\parbox[t]{0.8\textwidth}{
Shaded area denotes region of the $( \beta_{+} \,,  \beta_{-} )$ plane for which 
the ground-state Bethe roots obey
the string hypothesis for $p=2$, $N=2$, $\alpha_{-}=-1.818i$, 
$\alpha_{+}=2.959i$, $\theta_{-}=0.7i$, $\theta_{+}=0.6i$.}
	}
	\label{fig:p2beta}
\end{figure}

We emphasize that the result (\ref{XXZboundenergyeach}) has been
derived under the assumption that the Bethe roots for the ground state
obey the string hypothesis, which is true only for suitable
values of the boundary parameters. For example, the shaded areas in 
Figures \ref{fig:p2alpha} and  \ref{fig:p2beta} denote  
regions of parameter space for which the ground-state Bethe roots have 
the form described in Sections \ref{subsec:evensearoots} and
\ref{subsec:evenextraroots}. The $\alpha_{\pm}$  and $\beta_{\pm}$ parameters
are varied in the two figures, respectively.


\section{Odd $p$}\label{sec:odd}

\begin{figure}[htb]
	\centering
	\includegraphics[width=0.80\textwidth]{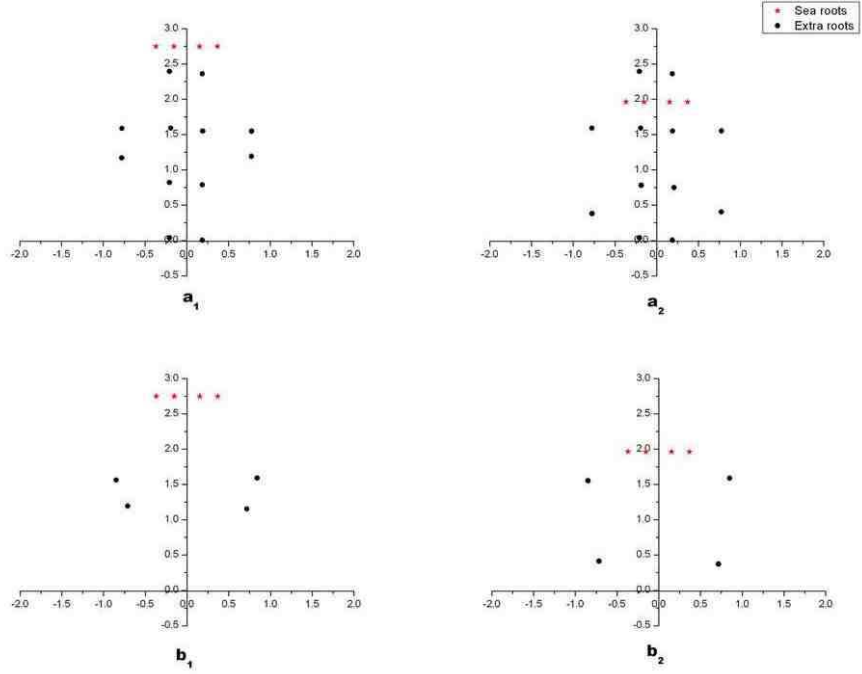}
	\caption[xxx]{\parbox[t]{0.8\textwidth}{
Ground-state Bethe roots for $p=3$, $N=4$, $\alpha_{-}=1.554i$, 
$\alpha_{+}=0.948i$, $\beta_{-}=-0.214$, $\beta_{+}=0.186$, 
$\theta_{-}=0.6i$, $\theta_{+}=0.7i$.}
	}
	\label{fig:p3}
\end{figure}

In this section, we consider the case where the bulk anisotropy parameter
assumes the values (\ref{etavalues}) with $p$ odd, i.e., 
$\eta ={i\pi\over 2}\,, {i\pi\over 4}\,, \ldots$.
As for the even $p$ case, for suitable values of 
the boundary parameters, the ground state Bethe roots 
$\{u_{k}^{(a_{j})}, u_{k}^{(b_{j})}\}$ have a regular 
pattern. An example with $p=3\,, N=4$ is shown in Figure \ref{fig:p3}.
As before, these roots can be categorized into
sea roots (the number of which depends on $N$) and extra roots (the
number of which depends on $p$) according to the following pattern 
which we adopt as our ``string hypothesis''.

\subsection{Sea roots 
$\{v_{k}^{\pm (a_{j})}, v_{k}^{\pm (b_{j})}\}$}\label{subsec:oddsearoots}

Sea roots of all $\{a_{j}(u) \,, b_{j}(u) \}$ functions for odd 
$p$ are given by
\be
v_{k}^{\pm (a_{j})} = v_{k}^{\pm (b_{j})} &=& \pm \tilde v_{k} 
+ \left({2 p + 3 - 2 j\over 2}\right)\eta\,, 
\quad  k = 1\,,\ldots\,, {N\over 2} \,,  \non\\
& & \qquad j = 1\,,\ldots \,, {p + 1\over 2} \,,  \label{seapodd}
\ee   
where $\tilde v_{k}$ are real and positive. In Figure \ref{fig:p3}, the sea 
roots are indicated with red stars.

As in the even $p$ case, the real parts ($\pm \tilde v_{k}$) are 
independent of $j$. This again provides simplification to the analysis. 
In contrast to the even $p$ case, now none of the functions 
$\{a_{j}(u) \,, b_{j}(u) \}$ has crossing symmetry. Hence, there are $N$ 
sea roots for all values of $j$.    

\subsection{Extra roots $\{w_{k}^{(a_{j},l)}, w_{k}^{(b_{j})}\}$} 
\label{subsec:oddextraroots}

We now describe the extra Bethe roots for odd $p$.  In Figure \ref{fig:p3}, the 
extra roots are indicated with black circles.
We start with the $p-1$ extra roots of the $b_{j}(u)$ functions:
\be
w_{k}^{\pm (b_{j})} &=& \pm \tilde w_{k} + \left(p - k\right)\eta\,,
\qquad k = 1\,,\ldots \,, p - 2 \,, \non\\
w_{p-1}^{\pm (b_{j})} &=& \pm \tilde w_{p-1} 
+ \left({p + 2 - 2 j\over 2}\right)\eta\,,
\qquad j = 1\,,\ldots \,, {p + 1\over2} \,. \label{extrabpodd}
\ee   
Similarly to the even $p$ case, the real parts of the extra roots 
are related to each other pairwise,
\be
\tilde w_{k}=\tilde w_{p - k-1}\,, \qquad k = 1\,,\ldots \,, {p-3\over 2} 
\,, \label{wkrltnodd}
\ee
so that only $\tilde w_{{p-1\over2}}$ remains unpaired. 

Similarly, the extra roots of the $a_{j}(u)$ functions are as follows,
\be
w_{k}^{\pm (a_{j},1)} &=& w_{k}^{\pm (b_{j})} = \pm \tilde w_{k} + \left(p - k\right)\eta\,,
\quad k = 1\,,\ldots \,, p - 2 \,, \non\\
 w_{p-1}^{\pm (a_{j},1)} &=& w_{p-1}^{\pm (b_{j})} =\pm \tilde w_{p-1} 
 + \left({p + 2 - 2 j\over 2}\right)\eta\,, \non \\
w_{k}^{\pm (a_{j},2)} &=& \pm \tilde w_{0} + \left(p + 1 - k\right)\eta\,,
\quad k = 1\,,\ldots \,, p + 1 \,, \qquad  j = 1\,,\ldots \,, {p + 
1\over 2} \,.
\ee  
As in the even $p$ case, the extra roots of the first type $\{
w_{k}^{\pm (a_{j},1)} \}$ coincide with the $b$ roots $\{ w_{k}^{\pm
(b_{j})} \}$. Moreover, the extra roots of the second type
$\{ w_{k}^{\pm (a_{j},2)} \}$ form a ``$(p + 1)$-string'', with real part 
$\tilde w_{0}$.

However, in contrast to the even $p$ case, 
some of the extra roots (namely, $w_{p-1}^{(a_{j},1)}$ 
and $w_{p-1}^{(b_{j})}$) depend on the value of $j$. Hence, as we shall 
see, these extra roots will not cancel from either the energy 
expression or the Bethe equations. Nevertheless, the contribution of these roots
to the boundary energy will ultimately cancel.

\subsection{Boundary energy}

As in the case of even $p$, we use the energy expression 
(\ref{energypgen}) and the string hypothesis to obtain (for $p \ge 3$)
\be
E &=& {1\over 2}\sinh \eta  \Bigg\{ \sum_{k=1}^{\frac{N}{2}} \Big[
\coth (v_{k}^{+(b_{j})}+(j-1)\eta) + \coth 
(v_{k}^{-(b_{j})}+(j-1)\eta) \non \\
&-&\coth (v_{k}^{+(b_{j-1})}+(j-1)\eta) - \coth (v_{k}^{-(b_{j-1})}+(j-1)\eta) \Big]\non \\
&+& \sum_{k=1}^{p-1} \Big[
\coth (w_{k}^{+(b_{j})}+(j-1)\eta) + \coth 
(w_{k}^{-(b_{j})}+(j-1)\eta) \non \\
&-&\coth (w_{k}^{+(b_{j-1})}+(j-1)\eta) - \coth (w_{k}^{-(b_{j-1})}+(j-1)\eta)\Big] 
\Bigg\} + E_{0}\,, \non \\
& & \qquad j= 2\,, \ldots \,, {p+1\over 2} \,.
\ee 
Recalling (\ref{seapodd}) and (\ref{extrabpodd}), this expression for
the energy reduces, independently of the value of $j$, to
\be
E &=& \sinh^{2}\eta \sum_{k=1}^{{N\over 2}}{1\over 
\sinh (\tilde v_{k} - {\eta\over2})
\sinh (\tilde v_{k} + {\eta\over2})}  
- {2\sinh^{2} \eta\over \cosh \eta + \cosh(2 \tilde w_{p-1})} + E_{0}\,,
\label{energyI2} 
\ee 
where $\tilde v_{k}\,, \tilde w_{p-1}  > 0$. As already anticipated, 
the expression for the energy depends on the extra root 
$\tilde w_{p-1}$ as well as on the sea roots.

Turning now to the Bethe Ansatz equations, following similar
arguments as for the even $p$ case, we find again that the $A$ 
normalization constants are all equal, and similarly for the
$B$'s, 
\be
A_{1}=A_{2}=\ldots=A_{{p+1\over 2}} \,, \qquad 
B_{1}=B_{2}=\ldots=B_{{p+1\over 2}} \,.
\ee 
Choosing $u_{l}^{(b_{j})}$ in (\ref{BAEpgenb})
to be a sea root $v_{l}^{+(b_{j})}$ for any $j \in \{2\,, \ldots\,,  
\frac{p+1}{2} \}$, we obtain
\be
\lefteqn{\left(
{\sinh(\tilde v_{l} + {\eta\over 2}) \over 
\sinh (\tilde v_{l} - {\eta\over 2})}\right)^{2N}
{\sinh(2\tilde v_{l} + \eta) \over 
\sinh (2\tilde v_{l} - \eta)}
{\sinh(\tilde v_{l} - {\eta\over 2} + \alpha_{-}) \over 
\sinh (\tilde v_{l} + {\eta\over 2} - \alpha_{-})}
{\cosh(\tilde v_{l} - {\eta\over 2} + \beta_{-}) \over 
\cosh (\tilde v_{l} + {\eta\over 2} - \beta_{-})}} \non \\
& & \times 
{\sinh(\tilde v_{l} - {\eta\over 2} + \alpha_{+}) \over 
\sinh (\tilde v_{l} + {\eta\over 2} - \alpha_{+})}
{\cosh(\tilde v_{l} - {\eta\over 2} + \beta_{+}) \over 
\cosh (\tilde v_{l} + {\eta\over 2} - \beta_{+})}
= - {\sinh(\tilde v_{l} - \tilde w_{p-1} - {p-1\over 2}\eta) \over 
\sinh (\tilde v_{l} - \tilde w_{p-1} + {p-1\over 2}\eta)}
{\sinh(\tilde v_{l} + \tilde w_{p-1} - {p-1\over 2}\eta) \over 
\sinh (\tilde v_{l} + \tilde w_{p-1} + {p-1\over 2}\eta)}\non \\
& & \times
\prod_{k=1}^{N\over 2} 
{\sinh(\tilde v_{l} - \tilde v_{k} + \eta) \over 
\sinh (\tilde v_{l} - \tilde v_{k} - \eta)}
{\sinh(\tilde v_{l} + \tilde v_{k} + \eta) \over 
\sinh (\tilde v_{l} + \tilde v_{k} - \eta)} \,,
\quad l = 1 \,, \cdots \,, {N\over 2} \,,\quad 
\tilde v_{k}\,, \tilde w_{p-1} > 0 \,.
\label{BAE2}
\ee 
In a compact form, this result can be written as 
\be 
e_{1}(\lambda_{l})^{2N+1}\  g_{1}(\lambda_{l})
{e_{2a_{-}-1}(\lambda_{l})\ e_{2a_{+}-1}(\lambda_{l})\over 
g_{1+2ib_{-}}(\lambda_{l})\ g_{1+2ib_{+}}(\lambda_{l})} &=& 
-\left[e_{p-1}(\lambda_{l}-\bar\lambda)\ 
e_{p-1}(\lambda_{l}+\bar\lambda)\right]^{-1}\non \\
\times\prod_{k=1}^{N\over 2}  
e_{2}(\lambda_{l}-\lambda_{k})\ 
e_{2}(\lambda_{l}+\lambda_{k})  \,, 
& & l = 1 \,, \cdots \,, {N\over 2} \,,
\label{BAEcompact2}
\ee
where $\tilde w_{p-1}=\mu\bar\lambda$.
The corresponding ground state counting function is given by
\be
\h(\lambda)&=&{1\over 2\pi}\Big\{ (2N+1) q_{1}(\lambda) + r_{1}(\lambda)
+ q_{2a_{-}-1}(\lambda) -  r_{1+2ib_{-}}(\lambda)
+ q_{2a_{+}-1}(\lambda) -  r_{1+2ib_{+}}(\lambda) \non \\
&+& q_{p-1}(\lambda- \bar\lambda) + q_{p-1}(\lambda+ \bar\lambda) 
-\sum_{k=1}^{N\over 2} \left[ q_{2}(\lambda - \lambda_{k}) +
q_{2}(\lambda + \lambda_{k})  \right] \Big\} \,.
\label{counting2}
\ee
Following similar procedure as before, we arrive at the root density for the 
ground state
\be
\rho(\lambda)&=& 2a_{1}(\lambda) 
-\int_{-\infty}^{\infty} d\lambda'\ a_{2}(\lambda - \lambda')\ \rho(\lambda')
+{1\over N}\Big[ a_{1}(\lambda) + b_{1}(\lambda) + a_{2}(\lambda) 
\label{counting3}\\
&+& a_{2a_{-}-1}(\lambda) -  b_{1+2ib_{-}}(\lambda)
+ a_{2a_{+}-1}(\lambda) -  b_{1+2ib_{+}}(\lambda) 
+ a_{p-1}(\lambda- \bar\lambda) + a_{p-1}(\lambda+ \bar\lambda)\Big] 
\,, \non 
\ee
where as before higher order corrections in $1/N$ are ignored when
passing from a sum to an integral. This yields 
\be
\rho(\lambda) = 2 s(\lambda) + {1\over N} R(\lambda) \,,
\label{rho2}
\ee
where now
\be
\hat R(\omega) &=& {1\over{\left(1+\hat a_{2}(\omega) \right)}} 
\Big\{ \hat a_{1}(\omega) + \hat b_{1}(\omega) +  \hat a_{2}(\omega)
  - \hat b_{1+2ib_{-}}(\omega) -  \hat b_{1+2ib_{+}}(\omega) \non \\
 &+&  \hat a_{2a_{-}-1}(\omega) +  \hat a_{2a_{+}-1}(\omega) 
 + 2 \cos (\bar\lambda\omega)\, \hat a_{p-1}(\omega)
   \Big\} \,.
\ee

The energy expression (\ref{energyI2}) yields, as $N\rightarrow\infty$,  
\be
E &=& -{2\pi \sin \mu\over \mu} \Big\{ 
\sum_{k=1}^{\frac{N}{2}} a_{1}(\lambda_{k}) 
+ b_{1}(\bar\lambda) \Big\} + E_{0}\non\\
&=& -{\pi \sin \mu\over \mu} \Big\{ 
\sum_{k=-\frac{N}{2}}^{\frac{N}{2}} a_{1}(\lambda_{k}) - a_{1}(0)
+ 2b_{1}(\bar\lambda) \Big\} + E_{0} \non\\
&=& -{\pi \sin \mu\over \mu} \Big\{ 
N\int_{-\infty}^{\infty}d\lambda\ a_{1}(\lambda)\ \rho(\lambda) - a_{1}(0) 
+ 2b_{1}(\bar\lambda) \Big\}
+ {1\over 2}(N-1) \cos \mu \non\\
&+& {1\over 2} \sin \mu \left( \cot \mu a_{-} + i\tanh \mu b_{-} +
\cot \mu a_{+} + i\tanh \mu b_{+} \right) \,.
\ee 
Substituting (\ref{rho2}) for the root density, we again obtain
\be
E = E_{bulk} + E_{boundary} \,,
\ee
where the bulk (order $N$) energy is again given by (\ref{bulkenergy}).
The boundary energy is now given by 
\be
E_{boundary} &=& - {\pi \sin \mu\over \mu} I
 -{1\over 2}\cos \mu + {1\over 2} \sin \mu \left( \cot \mu a_{-} + i\tanh \mu b_{-} +
\cot \mu a_{+} + i\tanh \mu b_{+} \right)\non\\
&-& {2\pi \sin \mu\over \mu} b_{1}(\bar\lambda) \,,
\label{boundenergyII}
\ee
where $I$ is now the integral 
\be
I &=& \int_{-\infty}^{\infty}d\lambda\ a_{1}(\lambda) \left[
R(\lambda) - \delta(\lambda) \right] 
= {1\over 2\pi}  \int_{-\infty}^{\infty} d\omega\ \hat a_{1}(\omega)
\left[ \hat R(\omega) - 1 \right] \non \\
&=& 
 {1\over 2\pi}  \int_{-\infty}^{\infty} d\omega\ \hat s(\omega)
\Big\{ \hat a_{1}(\omega) + \hat b_{1}(\omega)  - 
1 \non \\
& & -  \hat b_{1+2ib_{-}}(\omega) - \hat b_{1+2ib_{+}}(\omega) 
 + \hat a_{2a_{-}-1}(\omega) +  \hat a_{2a_{+}-1}(\omega) 
 + 2 \cos (\bar\lambda\omega)\, \hat a_{p-1}(\omega) 
 \Big\} \,. \label{integralII}
\ee
Using the fact that $\hat s(\omega) \hat a_{p-1}(\omega) = -\hat
b_{1}(\omega)$, we see that there is a perfect cancellation of the
last term in (\ref{boundenergyII}) which depends on the extra root
$\bar \lambda$.  Thus, as in the even $p$ case,
there is no contribution to the boundary energy
from extra roots.  Proceeding as before, we find that the energy
contribution from each boundary is again given by
(\ref{XXZboundenergyeach}), thus coinciding with previous results in
\cite{finitesize, MNS2}.  

\begin{figure}[htb]
	\centering
	\includegraphics[width=0.40\textwidth]{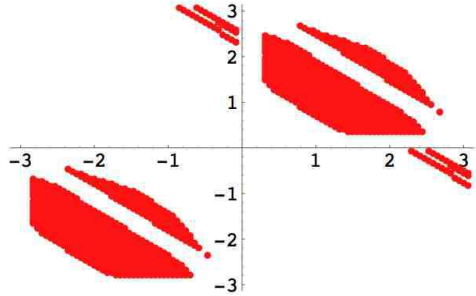}
	\caption[xxx]{\parbox[t]{0.8\textwidth}{
Shaded area denotes region of the $(\Im m\, \alpha_{+} \,, \Im m\,  
\alpha_{-} )$ plane for which the ground-state Bethe roots obey
the string hypothesis for $p=3$, $N=2$, $\beta_{-}=-0.85$, $\beta_{+}=0.9$, 
$\theta_{-}=0.6i$, $\theta_{+}=0.7i$.}
	}
	\label{fig:p3alpha}
\end{figure}

\begin{figure}[htb]
	\centering
	\includegraphics[width=0.40\textwidth]{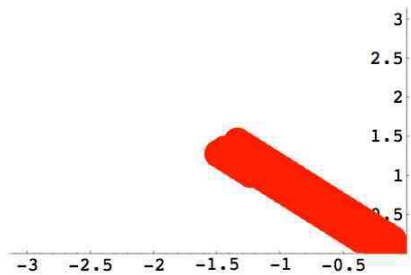}
	\caption[xxx]{\parbox[t]{0.8\textwidth}{
Shaded area denotes region of the $( \beta_{+} \,,  \beta_{-} )$ plane for which 
the ground-state Bethe roots obey
the string hypothesis for $p=3$, $N=2$, $\alpha_{-}=1.2i$, 
$\alpha_{+}=0.98i$, $\theta_{-}=0.7i$, $\theta_{+}=0.6i$.}
	}
	\label{fig:p3beta}
\end{figure}

As for even $p$, the derivation here is
based on the string hypothesis for the ground-state Bethe roots,
which is true only for suitable values of boundary
parameters. For example, the shaded areas in 
Figures \ref{fig:p3alpha} and  \ref{fig:p3beta} denote the 
regions of parameter space for which the ground-state Bethe roots have 
the form described in Sections \ref{subsec:oddsearoots} and
\ref{subsec:oddextraroots}. The $\alpha_{\pm}$  and $\beta_{\pm}$ parameters
are varied in the two figures, respectively.


\section{Discussion}\label{sec:conc}
       
We have studied the ground state of the general integrable open XXZ
spin-1/2 chain (\ref{Hamiltonian}) in the thermodynamic limit,
utilizing the solution we found recently in \cite{MNS}.  In contrast
to the earlier solution \cite{CLSW}-\cite{YNZ}, this solution does not
assume any restrictions or constraints among the boundary parameters.
However, the bulk parameter is restricted to values corresponding to
roots of unity (\ref{etavalues}).  The key to working with this
solution is formulating an appropriate string hypothesis, which leads
to a reduction of the Bethe Ansatz equations to a conventional form.
While the idea of using a string hypothesis to simplify the analysis
of Bethe equations is as old as the Bethe Ansatz itself, the
particular patterns appearing here are perhaps unparalleled in their
rich structure.

The boundary energy result (\ref{XXZboundenergyeach}) was obtained
previously \cite{finitesize} for bulk and boundary parameters that 
are unconstrained and constrained, respectively; and we have now 
obtained the same result for the reversed situation, namely, for 
bulk and boundary parameters that are constrained and unconstrained, 
respectively. Hence, this result presumably holds when both the bulk
and boundary parameters are unconstrained (within some suitable 
domains). Indeed, for the boundary sine-Gordon model \cite{GZ}, which is closely
related to the open XXZ chain, the expression \cite{BPT} for the boundary
energy is valid for general values of the bulk and boundary parameters.  In
view of the spectral equivalence between systems with diagonal and
nondiagonal boundary interactions noted in \cite{dG,Ni,Baj}, it may be
interesting to try to relate our boundary energy result with the
corresponding result \cite{HQB} for diagonal boundary interactions.

Having demonstrated the practicality of this solution, we now
expect that it should be possible to use a similar approach
to analyze further properties of the model, such as the Casimir energy
(order $1/N$ correction to the ground state energy), and bulk and 
boundary excited states.

There is an evident redundancy in the solution which we have used
here: there are many equivalent expressions for the energy (see, e.g.,
(\ref{energypgen}), (\ref{energyspecial})), and we find that the Bethe
Ansatz equations (\ref{BAEpgena}), (\ref{BAEpgenb}) all become
equivalent upon imposing the string hypothesis.  Moreover, while there
are various ``extra'' Bethe roots describing the ground state, they
ultimately do not contribute to the boundary energy.  All of this
suggests that it may be possible to find a simpler and more economical
solution of the model involving fewer $Q$ functions. Ideally, one 
would like to find a solution for which neither bulk nor boundary 
parameters are constrained.       

\section*{Acknowledgments}

This work was supported in part by the National Science Foundation
under Grants PHY-0244261 and PHY-0554821.

\begin{appendix}

\section{Appendix}\label{sec:expressions}

We list here explicit expressions for the function $f_{1}(u)$ and the 
coefficients $\mu_{k}$ appearing in the
text. We remind the reader that we assume throughout that $N$ is even.
 
For even values of $p$, the function $f_{1}(u)$ appearing in the 
Bethe Ansatz equations (\ref{BAEpgena1}) - (\ref{BAEpgenb}),  (\ref{BAEnorm}) 
is given by
\be
f_{1}(u) &=& - 2^{3-2 p} \Big( \non \\
& & \hspace{-0.2in}
\sinh \left( (p+1) \alpha_{-} \right)\cosh \left( (p+1) \beta_{-} \right)
\sinh \left( (p+1) \alpha_{+} \right)\cosh \left( (p+1) \beta_{+} \right)
\cosh^{2} \left( (p+1)u \right) \non \\
&-&
\cosh \left( (p+1) \alpha_{-} \right)\sinh \left( (p+1) \beta_{-} \right)
\cosh \left( (p+1) \alpha_{+} \right)\sinh \left( (p+1) \beta_{+} \right)
\sinh^{2} \left( (p+1)u \right) \non \\
&-&
 \cosh \left( (p+1)(\theta_{-}-\theta_{+}) \right)
\sinh^{2} \left( (p+1)u \right) \cosh^{2} \left( (p+1)u \right) 
\Big) \,.
\label{f1even}
\ee
For odd values of $p$,
\be
f_{1}(u) &=& -2^{3-2 p} \Big( \non \\
& & \hspace{-0.2in}
\cosh \left( (p+1) \alpha_{-} \right)\cosh \left( (p+1) \beta_{-} \right)
\cosh \left( (p+1) \alpha_{+} \right)\cosh \left( (p+1) \beta_{+} \right)
\sinh^{2} \left( (p+1)u \right) \non \\
&-&
\sinh \left( (p+1) \alpha_{-} \right)\sinh \left( (p+1) \beta_{-} \right)
\sinh \left( (p+1) \alpha_{+} \right)\sinh \left( (p+1) \beta_{+} \right)
\cosh^{2} \left( (p+1)u \right) \non \\
&+&
 \cosh \left( (p+1)(\theta_{-}-\theta_{+}) \right)
\sinh^{2} \left( (p+1)u \right) \cosh^{2} \left( (p+1)u \right) 
\Big) \,. \label{f1odd}
\ee 
For both even and odd values of $p$, these functions have the properties
\be
f_{1}(u + \eta) = f_{1}(u) \,, \qquad f_{1}(-u)=f_{1}(u).
\label{fident}
\ee

The coefficients $\mu_{k}$ appearing in the function $Y(u)$ (\ref{Y})
are given as follows for even (upper sign) and odd (lower sign) values of $p$.
\be
\mu_{0} &=& 2^{-4p} \Bigg\{ -1 - 
\cosh^{2}((p+1)(\theta_{-}-\theta_{+})) \non \\ 
&-& \cosh(2(p+1)\alpha_{-}) \cosh(2(p+1)\alpha_{+}) 
\mp \cosh(2(p+1)\alpha_{-}) \cosh(2(p+1)\beta_{-}) \non \\ 
&\mp& \cosh(2(p+1)\alpha_{+}) \cosh(2(p+1)\beta_{-}) 
\mp \cosh(2(p+1)\alpha_{-}) \cosh(2(p+1)\beta_{+}) \non \\ 
&\mp& \cosh(2(p+1)\alpha_{+}) \cosh(2(p+1)\beta_{+}) 
- \cosh(2(p+1)\beta_{-})  \cosh(2(p+1)\beta_{+}) \non \\ 
&+& \Big[ \cosh((p+1)(\alpha_{-}+\alpha_{+})) 
\cosh((p+1)(\beta_{-}-\beta_{+})) \non \\
&\pm& \cosh((p+1)(\alpha_{-}-\alpha_{+})) \cosh((p+1)(\beta_{-}+\beta_{+})) \Big]^{2}\non \\ 
&\mp& 2 \cosh((p+1)(\theta_{-}-\theta_{+})) \Big[
\cosh((p+1)(\alpha_{-}-\alpha_{+})) \cosh((p+1)(\beta_{-}-\beta_{+})) \non \\ 
&\pm& \cosh((p+1)(\alpha_{-}+\alpha_{+})) \cosh((p+1)(\beta_{-}+\beta_{+})) 
\Big] \Bigg\}
\,, \non \\ 
\mu_{1} &=& 2^{1-4p} \Bigg\{ 
\cosh((p+1)(\alpha_{-}\mp\alpha_{+})) 
  \Big[ \cosh((p+1)(\alpha_{-}\pm\alpha_{+})) \non \\
  &+& 
\cosh((p+1)(\beta_{-}\pm\beta_{+}))\cosh((p+1)(\theta_{-}-\theta_{+})) 
\Big] \non \\
&\mp& \cosh((p+1)(\beta_{-}\mp\beta_{+}))
\Big[ \cosh((p+1)(\beta_{-}\pm\beta_{+}))  \non \\
 &+& 
\cosh((p+1)(\alpha_{-}\pm\alpha_{+}))\cosh((p+1)(\theta_{-}-\theta_{+})) \Big]
\Bigg\} \,, \non \\ 
\mu_{2} &=& 2^{-4p} \sinh^{2}((p+1)(\theta_{-}-\theta_{+})) \,. 
\ee 

\end{appendix}

\end{document}